\documentclass[]{spie}  

\usepackage{amsmath,amsfonts,amssymb}
\usepackage{multirow}
\usepackage{graphicx}
\usepackage[colorlinks=true, allcolors=blue]{hyperref}

\title{MATISSE: specifications and expected performances}

\author[a]{A. Matter}
\author[a]{S. Lagarde}
\author[a]{R. G. Petrov}
\author[a]{P. Berio}
\author[a]{S. Robbe-Dubois}
\author[a]{B. Lopez}
\author[a]{P. Antonelli}
\author[a]{F. Allouche}
\author[a]{P. Cruzalebes}
\author[a]{F. Millour}
\author[b]{G. Bazin}
\author[c]{L. Bourg\`es}

\affil[a]{Laboratoire Lagrange, Universit\'e C\^ote d'Azur, Observatoire de la C\^ote d'Azur, CNRS, Boulevard de l'Observatoire, CS 34229, 06304 Nice, France.}
\affil[b]{European Southern Observatory.}
\affil[c]{Universit\'e de Grenoble Alpes, CNRS, IPAG, Grenoble, France.}

\authorinfo{Further author information: (Send correspondence to A. Matter)\\A. Matter: E-mail: Alexis.Matter@oca.eu, Telephone: +33 (0)4 92 00 19 79}

\begin{document} 
\maketitle

\begin{abstract}
MATISSE (Multi AperTure mid-Infrared SpectroScopic Experiment) is the next generation spectro-interferometer at the European Southern Observatory VLTI operating in the spectral bands {\itshape L, M} and {\itshape N}, and combining four beams from the unit and auxiliary telescopes. MATISSE is now fully integrated at the Observatoire de la C\^ote d'Azur in Nice (France), and has entered very recently its testing phase in laboratory. This paper summarizes the equations describing the MATISSE signal and the associated sources of noise. The specifications and the expected performances of the instrument are then evaluated taking into account the current characteristics of the instrument and the VLTI infrastructure, including transmission and contrast degradation budgets. In addition, we present the different MATISSE simulation tools that will be made available to the future users.  
\end{abstract}


\section{INTRODUCTION}
\label{sec:intro}  
The Multi AperTure mid-Infrared SpectroScopic Experiment (MATISSE) (Matter et al, 2016 this volume) is the spectro-interferometer under construction for the VLTI
combining four beams from the telescopes UTs or ATs. Covering the {\itshape L, M},
and {\itshape N} bands, this second generation instrument will address several f­undamental research topics in astrophysics, including the inner regions of disks around young stars where planets form and evolve, the surface structure and mass loss of stars at different evolutionary stages, and the environment of black holes in active galactic nuclei (AGN).
The project successfully passed the Final Design Review (FDR) in April 2012. This resulted in the validation of the
instrumental study and the compliance of the estimated performance with the expected scientific specifications, and in the
acceptance of ESO for the MATISSE consortium to proceed to the manufacturing, integration and tests of the instrument. MATISSE is now fully integrated at the Observatoire de la C\^ote d'Azur in Nice (France), and has entered very recently its testing phase in laboratory. The purpose of this paper is to present a short overview on the current noise model of MATISSE and the estimated performances (limiting
magnitude, visibility and phase accuracy), as previously presented in detail in Lagarde et al. (2012)\cite{2012SPIE.8445E..2JL}. We also describe the upcoming VLTI upgrades that will have an impact on the MATISSE performances, and the MATISSE simulation tools. After a short introduction on the instrument concept and on the
available observing modes in Sect. 2, Sect. 3 explains the different techniques to get rid of the thermal background
effects and compares the so-expected results with the requirement. Section 4 gives the noise model of MATISSE and the different formula necessary to
calculate the performance of MATISSE in terms of coherent flux, photometry, visibility, closure and differential phases.
Section 5 provides a summary on the estimation of the expected instrumental visibility on an unresolved source and of the number of
photons received on the detector. The formula of Sect. 4 and the estimated values of Sect. 5 are the necessary inputs
to calculate the performance presented in Sect. 6. Section 7 presents the upcoming upgrades on the VLTI infrastructure that will improve the MATISSE performances. Finally, in Sect. 8, we present the different MATISSE simulation tools that will be made available to the future users. 

\section{Instrument concept and observing modes}
\label{sec:concept}
MATISSE is a four-beam experiment with a multi-axial global combination. In addition to the interferometric image
containing the pattern with six fringe periods, the observation of the images of each individual beam is made possible for
photometric calibration. In this case, five images are formed onto the detector. The interferometric beam and the four
photometric beams receive respectively 2/3 and 1/3 of the incoming flux. The interferogram is dispersed in the spectral
direction. Two spectral resolutions are provided in {\itshape N} band, $R = 30$ and $R=220$, and four in {\itshape L} and {\itshape M} bands, $R = 30$, $R=500$, $R=1000$ and $R=3500-5000$. The size of the interferometric channel is larger than the photometric ones in order to optimize the sampling of
the 6 fringe patterns. The beam combination is made by the camera optics. At this level, the beam configuration is non
redundant (separation $B_{\rm ij}$ between beams $i$ and $j$ respectively equal to $3D$, $9D$ and $6D$ where $D$ is the beam diameter) in
order to avoid crosstalk between the fringe peaks in the Fourier space.
The Fourier transform of each spectral column of the interferometric image is thus composed of six fringe peaks centered
at different frequencies $B_{\rm ij}/\lambda$ ($3D/\lambda$, $6D/\lambda$, $9D/\lambda$, $12D/\lambda$, $15D/\lambda$, $18D/\lambda$), and a low frequency peak that contains the object photometry and the thermal background collected by the four telescopes. The analytical expressions of the Fourier transform $I(u)$ of the interferometric channel, and of the Fourier transform $P_{\rm i}(u)$
of each photometric channel $i$ are given by:
\begin{align}
I(u) &= M_{\rm b}(u).\sum_{i=1}^4n^I_{\rm b_{\rm i}}+M(u).\sum^4_{i=1}n^I_{\rm i}+\sum^4_{i=1}\sum^4_{j=2}M(u-u_{\rm ij})\sqrt{n^I_{\rm i}.n^I_{\rm j}}V_{\rm ij}, \\
P_{\rm i}(u) &=M_{\rm b}(u).n^P_{\rm b_{\rm i}}+M(u).n_{\rm i}^P,
\end{align}
where $i$ and $j$ are the index of beams, $n^I_{\rm b_{\rm i}}$ is the number of photons produced by the thermal background for beam $i$ in the interferometric channel, $n^P_{\rm b_{\rm i}}$ is the number of photons produced by the thermal background in the photometric channel $i$, $n^I_{\rm i}$ and $n^I_{\rm j}$ are the numbers of photons produced by the observed object for each beam in the interferometric channel, $n_{\rm i}^P$ is the number of photons produced by the observed object in the photometric channel $i$, and $V_{\rm ij}$ is the complex visibility.
$M_{\rm b}(u)$ is the frequency peak function of the thermal background in the Fourier space, $M(u)$ is the low frequency peak of the interferometer transfer function, $M(u-u_{\rm ij})$ is the fringe peak of the interferometer transfer function at the spatial frequency $u_{\rm ij}$ ($u_{\rm ij} = B_{\rm ij}/\lambda)$. At the zero frequency, $M(0)=M_{\rm b}(0)=1$.
Analogously to the VLTI instrument MIDI \cite{2003Ap&SS.286...73L}, MATISSE will operate primarily in two modes:
\begin{enumerate}
\item "SiPhot" (Simultaneous Photometry) mode: the SiPhot mode consists in measuring photometry simultaneously to the fringes acquisition. Chopping, a fast switch between object and sky observations, is used to remove thermal background from the photometric measurements. In this way, the average source photometry can be measured and then used to extract the
visibility from the coherent flux (measured at the same time).
\item "HighSens" (High Sensitivity) mode: the HighSens mode has no simultaneous photometry and all photons are collected in the interferometric beam. This maximizes the sensitivity and also the SNR on the coherent flux measurements and then the differential and closure phases. Chopping is optional in this mode, as well as subsequent photometry measurements.
\end{enumerate}
When the telescopes are pointed on sky and the chopping is used, only the thermal background is recorded on the detector. Eq.1 reduces so to the first term:
\begin{equation}
I_{\rm S}(u)=M_{\rm b}(u).\sum_{\rm i=1}^{4}n_{\rm b_{\rm i}}^I,
\end{equation}
and Eq.2 to the following :
\begin{equation}
I_{\rm S}(u)=M_{\rm b}(u).n_{\rm b_{\rm i}}^P.
\end{equation}
In order to measure the coherent fluxes with a good accuracy, the design of MATISSE is based on the use of spatial filters, including image and pupil stops. A beam commutation can also be made in order to reduce the effect of the instrumental defects on the measurement of the closure phase and the differential phase. This is performed by a specific system, the Beam Commuting Device (BCD).
\section{Background subtraction}
\label{sec:background}
Because the thermal background at the longest wavelengths (mainly in {\itshape N} band) is variable and far exceeds the target coherent flux, it is important to limit cross-talk between the low frequency peak and the high frequency peaks in the Fourier space. We have to damp the contamination of the fringe peaks by the background peak at a level below the photon noise limit, which is the square root of the thermal background emission. In {\itshape N} band, estimates based on MIDI data indicate that this requires an attenuation factor of about 5.10$^{-7}$. Two methods are used in MATISSE to ensure this result with a large margin : spatial modulation, like in the VLTI near-infrared spectrometer AMBER\cite{2007A&A...464....1P}, combined with temporal modulation like in MIDI. In Sect. 3.1 and 3.2, we evaluate the efficiency of these methods. In addition, to measure the visibility we also need to extract the source photometry, which consists in separating the stellar flux from the skybackground emission using chopping as described in Sect. 3.3.

\subsection{Spatial modulation}
\begin{figure} [t]
  \begin{center}
  \begin{tabular}{c} 
   \includegraphics[scale=1]{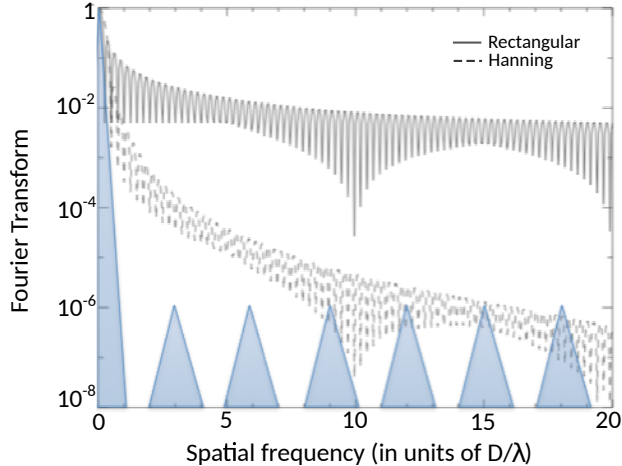}
   \end{tabular}
   \end{center}
   \caption[example] 
   { \label{fig:TF} 
Normalized Fourier transform of the digital window for a rectangular (up) and a Hanning window (down).}
   \end{figure}

Due to the geometry of the output pupil of diameter $D$, the size of the low frequency peak and the fringe peaks would be strictly reduced to $2D/\lambda$ if the detector window was infinitely wide. In practice, the theoretical spectral density of the interferogram is convolved by the Fourier transform of the finite window. Figure \ref{fig:TF} presents the Fourier transform of two digital windows, a flat rectangular window and a window with a Hanning apodizing profile, both with a size of
$4\lambda/D$. The effect is that the background term at low frequency is rejected to high frequencies and therefore contaminates the fringes peaks of the spectral density.
A rejection factor $\gamma_{\rm spat}$ can be estimated by integrating the Fourier transform of the digital window over the support of the fringe peaks. Table \ref{tab:rejection} gives $\gamma_{\rm spat}$ for each fringe peak for the two digital windows.

\begin{table}[ht]
\caption{Rejection factor $\gamma_{\rm spat}$ due to the spatial modulation, and shown for the six fringe peaks and the two digital windows (rectangular and Hanning). 
} 
\label{tab:rejection}
\begin{center}       
\begin{tabular}{ccccccc} 
\hline
&{\small $\boldsymbol{3D/\lambda}$} & {\small $\boldsymbol{6D/\lambda}$}&{\small $\boldsymbol{9D/\lambda}$}&{\small $\boldsymbol{12D/\lambda}$}&{\small $\boldsymbol{15D/\lambda}$}&{\small $\boldsymbol{18D/\lambda}$} \\
\hline
\textbf{Rectangular}&{\small $7.2\:10^{-3}$}&{\small $2.5\:10^{-3}$}&{\small $1.7\:10^{-3}$}&{\small $1.4\:10^{-3}$}&{\small $1.3\:10^{-3}$}&{\small $1.2\:10^{-3}$}\\
\hline
\textbf{Hanning}&{\small $8.8 \:0^{-5}$}&{\small $4.7\:10^{-6}$}&{\small $1.1\:10^{-6}$}&{\small $4.1\:10^{-7}$}&{\small $1.9\:10^{-7}$}&{\small $9.8\:10^{-8}$}\\
\hline 
\end{tabular}
\end{center}
\end{table}
It appears that the Hanning window provides a better rejection factor than the rectangular window. Nevertheless, $\gamma_{\rm spat}$ does not reach the required value of $\sim 5.10^{-7}$ for all the peaks. This indicates that the technique of spatial modulation is not sufficient alone to suppress the thermal background effect to the required level.

\subsection{Temporal modulation}
\begin{figure} [t]
  \begin{center}
  \begin{tabular}{c} 
   \includegraphics[scale=0.6]{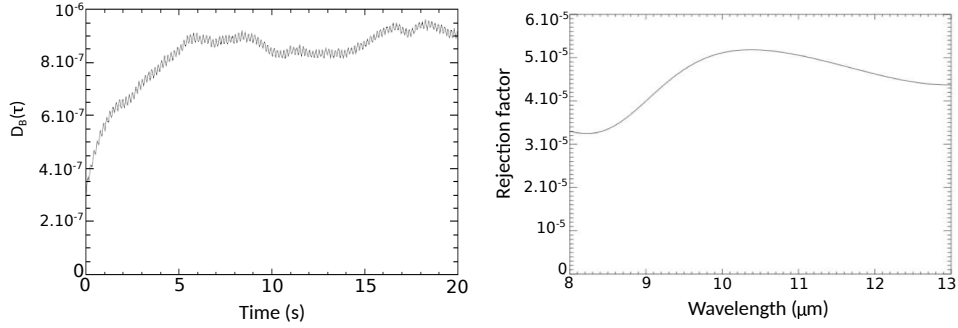}
   \end{tabular}
   \end{center}
   \caption[example] 
   { \label{fig:rejection} 
Left: Thermal background structure function obtained with MIDI data. Right: Rejection rate $\gamma_{\rm temp}$ after a temporal modulation of 10 steps of 1.3~$\mu$m and 30 ms each, in {\itshape N} band.}
   \end{figure}

The optical path difference (OPD) modulation consists in modulating each fringe signal by applying an artificial periodic sequence during a specific time called the modulation cycle. Without fringe tracking, a modulation cycle cannot last longer than the atmospheric coherence time. In {\itshape N} band, a typical modulation is applied during the 300 ms atmospheric coherence time, and 
consists in 10 steps of $\lambda_{\rm max}/10$ lasting 30~ms each. The resulting 10 signals are then numerically re-phased and summed. The lower frequency peak, which is not affected by the OPD modulation, is eliminated and only the high frequency peaks remain. The interferometric signal is sensitive to background variations during one modulation cycle. This remaining background is at the origin of the rejection factor associated with temporal modulation. This rejection rate can be estimated from the structure function of the thermal background. The structure function is defined as :
\begin{equation}
D_{\rm B}(\tau)=<[n_{\rm b}(t)-n_{\rm b}(t+\tau)]^2>.
\end{equation}
We have used an empirical law deduced from MIDI measurements in {\itshape N} band and written as :
\begin{equation}
D_{\rm B}(\tau)=2\left(\frac{n_{\rm b}}{1000}\right)^2\:\left(1-\exp{-0.693\tau}\right).
\end{equation}
Figure \ref{fig:rejection} (left) shows an example of the structure function in {\itshape N} band (data obtained with MIDI on the UTs). This function is normalized by the average number of photons due to the thermal background. The result of an OPD modulation simulation is given in figure \ref{fig:rejection} (right). We see that the temporal modulation rejection rate is $\gamma_{\rm temp}< 5.10^{-5}$. In the worst case (for the first peak and a rectangular window, see Table 1, the combination of spatial and temporal modulation reduces the contamination by the low frequency
peak by a factor $\gamma=\gamma_{\rm spat}.\gamma_{\rm temp}=4.10^{-7}$. This fits with the requirement and shows that this combination ensures a coherent flux estimate independent from the worst possible background fluctuations, while temporal or spatial modulation alone are not sufficient.
\subsection{Chopping}
The chopping system of the VLTI allows observing successively the object plus sky, and the sky alone. In MATISSE, chopping is used to extract, in the photometric beams, the source photometry from the background level. It is mandatory for accurate visibility measurements (SiPhot mode”). However, with chopping, the observations on sky and on the the target are not simultaneous. The corresponding error on the measurement is given by the structure function of the background (see Eq.5). Therefore, chopping yields to an attenuation of the background set by:
\begin{equation}
\gamma_{\rm chop}=\frac{\sqrt{D_B(2/f_{\rm chop})}}{\bar{n}_{\rm b}}=\frac{\sqrt{1-\exp{-1.39\:f_{\rm chop}}}}{707},
\end{equation}
where $\bar{n}_{\rm b}$ is the background flux per photometric beam, and $f_{\rm chop}$ is the chopping frequency. Table~\ref{tab:chop_rejection} shows the chopping rejection factor for different chopping frequencies.
\begin{table}[ht]
\caption{Rejection factor of the thermal background for different chopping frequencies). 
} 
\label{tab:chop_rejection}
\begin{center}       
\begin{tabular}{lcccc} 
\hline
{\small \textbf{Chopping frequency} $f_{\rm chop}$}~{\footnotesize (Hz)} & {\small 0.5} & {\small 0.8} &{\small 1}&{\small 3} \\
\hline
{\small \textbf{Background rejection} $\gamma_{\rm chop}$}&{\small $10^{-3}$}&{\small $1.2 10^{-3}$}&{\small $1.2 10^{-3}$}&{\small $1.4 10^{-3}$}\\
\hline 
\end{tabular}
\end{center}
\end{table}
It appears that chopping alone is not sufficient to measure the coherent flux. So we need spatial and temporal modulation. Nevertheless, chopping is the only way to extract the photometry.

\section{Signal to noise ratio}
This section gives the different formula necessary to calculate the performance of MATISSE in terms of coherent flux, photometry, visibility, closure and differential phase, with respect to the photon statistics received at the detector, the detector readout noise, and the residual background fluctuations. All the measurement estimators are deduced from the equations of Sect. 2. They will be used in Sect. 6 to estimate the performance of the instrument.

\subsection{Coherent flux SNR}
Assuming $M_{\rm b}(u_{\rm ij})$, $M(u_{\rm ij})$ and $M(u-u_{\rm ij}$, $u \neq u_{\rm ij}$) are small compared to noise (by using a sky information and/or an OPD modulation), the thermal background independent coherent flux $C_{\rm ij} =(n_i\:n_j)^{1/2}\:V_{\rm ij}$ can be extracted from the following estimator: $E(C_{\rm ij})= I(u_{\rm ij})$. The variance of this estimator is:
\begin{equation}
\sigma_{\rm C}^2=4n_{\rm b}^{\rm I}+4n^{\rm I}+n_{\rm P}'{\rm RON}^2,
\end{equation}
where $n^{\rm I}$ is the number of 'science' photons per beam in the interferometric channel, $n^{\rm I}_{\rm b}$ is the number of thermal background photons per beam in the interferometric channel and $n_{\rm P}'$ is the number of pixels used to analyze the signal in the interferometric channel. In the case of a finite window, a part of the low frequency peak is aliased to high frequency. Therefore, the coherent flux estimation is degraded by the contribution of the thermal background. So the signal to noise ratio (SNR) of the coherent flux per frame is:
\begin{equation}
SNR_{\rm C}=\frac{n^{\rm I}V}{4n_{\rm b}^{\rm I}+4n^{\rm I}+n_{\rm P}'{\rm RON}^2+16\gamma^2n_{\rm b}^2}
\end{equation}
$\gamma$ being the gobal rejection rate after spatial and temporal modulation (see Sect. 3.2)

\subsection{Photometry SNR}
The photometry estimation requires the estimation of the sky background, for instance with chopping. Let’s assume a chopping ratio equal to 0.5 (identical time for the observation of target+sky and sky alone). The photometry can be extracted from the following estimator: $E(n_{\rm i})=P_{\rm i}-S_{\rm i}$, $P_{\rm i}$ is the photometry channel of beam $i$ (during observation on target+sky) and $S_{\rm i}$ is the photometric channel of beam $i$ (during sky pointing). The variance of this estimator is:
\begin{equation}
\sigma_{\rm n}^2=n^{\rm P}+2n_{\rm b}^{\rm P}+2n_{\rm P}''{\rm RON}^2,
\end{equation}
where $n^{\rm P}$ is the number of 'science' photons per beam in the photometric channel, $n_{\rm b}^{\rm P}$ is the number of thermal background photons per beam in the photometric channel and $n_{\rm P}''$ is the number of pixels used to analyze the signal in the photometric channel. In the case of temporal fluctuations, an additional error in the sky estimation has to be taken into account, as the sky is not observed simultaneously to the target. As written in Sect. 3.3, this error can be estimated from the structure function of the background fluctuations $D_{\rm B}(\tau)$ (see Eq. 6). The SNR of the photometry per frame is then:
\begin{equation}
SNR_{\rm n}=\frac{n^{\rm P}}{2n_{\rm b}^{\rm P}+n^{\rm P}+2n_{\rm P}''{\rm RON}^2+D_{\rm B}\left(1/f_{\rm chop}\right)}
\end{equation}

\subsection{Visibility SNR}
The estimator of the visibility is $V=C/n$ where $C$ is the coherent flux and $n$ the photometry.
The error on the visibility is:
\begin{equation}
\sigma_{\rm V}^2=\left(\frac{1}{n}\right)^2\sigma_{\rm C}^2+\left(\frac{C}{n^2}\right)^2\sigma_{\rm n}^2=V^2\left[\frac{\sigma_{\rm C}^2}{C^2}+\frac{\sigma_{\rm n}^2}{n^2} \right] 
\end{equation}
So the SNR of the visibility estimator is: 
\begin{equation}
SNR_{\rm V}=\frac{1}{\sqrt{\frac{1}{SNR_{\rm C}^2}+\frac{1}{SNR_{\rm n}^2}}}
\end{equation}

\subsection{Error on the closure phase}
Following Petrov et al. (2007)\cite{2007A&A...464....1P} , the error on the closure phase can be approximated by :
\begin{equation}
\sigma_{\psi}^2\approx\frac{1}{2}\left(\frac{1}{SNR_{\rm C_1}^2}+\frac{1}{SNR_{\rm C_2}^2}+\frac{1}{SNR_{\rm C_3}^2} \right),
\end{equation}
where the index $i=1$, 2 or 3 refers to one of the three baselines involved in the closure phase. This approximation applies for bright enough images, for which $\sigma_{\psi}<1$~rad per frame. In fact, this error expression is derived assuming that the three phases are independent.
\subsection{Error on the differential phase}
Following Petrov et al. (2007)\cite{2007A&A...464....1P} , the error on the differential phase $\phi=\phi_{\rm i}-\phi_{\rm REF}$ can be approximated by :
\begin{equation}
\sigma_{\phi}^2\approx \sigma_{\phi_{\rm i}}^2+\sigma_{\phi_{\rm REF}}^2 \approx \frac{1}{2}\left(\frac{1}{SNR_{\rm C_{\rm i}}^2}+\frac{1}{SNR_{\rm REF}^2} \right)
\end{equation}
$\phi_{\rm i}$ is the phase for one spectral channel, and $\phi_{\rm REF}$ is the mean phase of the entire spectrum. In this case, $SNR_{\rm C_{\rm i }}<<SNR_{\rm REF}$. So the final expression is:
\begin{equation}
\sigma_{\phi}^2\approx\frac{1}{2SNR_{\rm C_{\rm i}}^2}
\end{equation}

\section{Error budget}
This section summarizes the estimates of the expected instrumental visibility V on an unresolved source (Sect. 5.1), the number of photons received on the detector (Sect. 5.2), and the detector parameters (Sect. 5.3). These parameters are necessary to calculate, using the formulas in Sect. 4, the performance presented in Sect. 6. The complete details on the error budget can be found in Lagarde et al. (2012)\cite{2012SPIE.8445E..2JL}.

\subsection{Visibility on unresolved source}
The instrument+atmosphere contrast (visibility on an unresolved source) is given by $V = V_{\rm I}V_{\rm F}V_{\rm C}V_{\rm J}$ where:
\begin{itemize}
\item $V_{\rm I}$ describes the instrumental contrast obtained on a perfect unresolved source feeding MATISSE. It depends on the instrumental characteristics.
\item $V_{\rm F}$ describes the contrast obtained after the spatial filter depending on the spatial filter characteristics, on the Strehl ratio (the atmosphere WFE), the image and pupil lateral motions at the spatial filter level.
\item $V_{\rm C}$ describes the contrast resulting from the chromatic OPD in the VLTI tunnel.
\item $V_{\rm J}$ describes the contrast resulting from the OPD shift jitter during a frame time or during a temporal modulation cycle. Two modes are considered here: 'blind mode', where OPD stability is assumed for MATISSE, and 'Fringe tracker', where an external fringe tracker with similar performance as FINITO is considered.
\end{itemize}
The error budget for these four terms is detailed in Lagarde et al. (2012)\cite{2012SPIE.8445E..2JL}. The results on the global contrast are given in Table~\ref{tab:instru_vis}
\begin{table}[ht]
\caption{Estimated global contrast $V$.
} 
\label{tab:instru_vis}
\begin{center}       
\begin{tabular}{lcccccc} 
\hline
\multirow{2}{*}{\textbf{Contrast}}&\multicolumn{2}{c}{\textbf{{\itshape L} band}} & \multicolumn{2}{c}{\textbf{{\itshape M} band}}& \multicolumn{2}{c}{\textbf{{\itshape N} band}}\\
\cline{2-7}
&\textbf{AT}&\textbf{UT}&\textbf{AT}&\textbf{UT}&\textbf{AT}&\textbf{UT}\\
\hline
{\small $V_{\rm I}$ (Low Spectral Res.)}&{\small 0.74}&{\small 0.74}&{\small 0.81}&{\small 0.81}&{\small 0.66}&{\small 0.66} \\
\hline
{\small $V_{\rm I}$ (Med. \& High Spectral Res.)}&{\small 0.76}&{\small 0.76}&{\small 0.84}&{\small 0.84}&{\small 0.68}&{\small 0.68} \\
\hline
{\small $V_{\rm F}$} &{\small 0.98}&{\small 0.97}&{\small 0.99}&{\small 0.99}&{\small 0.99}&{\small 0.99} \\
\hline
{\small $V_{\rm I}$ (Low Spectral Res.)}&{\small 0.92}&{\small 0.92}&{\small 0.95}&{\small 0.95}&{\small 0.96}&{\small 0.96} \\
\hline
\multicolumn{7}{c}{\textbf{BLIND MODE}}\\
\hline
{\small $V_{\rm J}$} &{\small 0.92}&{\small 0.75}&{\small 0.95}&{\small 0.84}&{\small 0.71}&{\small 0.62} \\
\hline
{\small $\boldsymbol{V}$} {\footnotesize \textbf{Low Spectral Res.}} &{\small \textbf{0.61}}&{\small \textbf{0.50}}&{\small \textbf{0.72}}&{\small \textbf{0.64}}&{\small \textbf{0.45}}&{\small \textbf{0.39}} \\
\hline
{\small $\boldsymbol{V}$} {\footnotesize \textbf{Med. \& High Spectral Res.}} &{\small \textbf{0.69}}&{\small \textbf{0.55}}&{\small \textbf{0.79}}&{\small \textbf{0.70}}&{\small \textbf{0.48}}&{\small \textbf{0.42}} \\
\hline
\multicolumn{7}{c}{\textbf{FRINGE TRACKER}}\\
\hline
{\small $V_{\rm J}$} &{\small 0.96}&{\small 0.72}&{\small 0.97}&{\small 0.82}&{\small 0.99}&{\small 0.96} \\
\hline
{\small $\boldsymbol{V}$} {\footnotesize \textbf{Low Spectral Res.}} &{\small \textbf{0.64}}&{\small \textbf{0.48}}&{\small \textbf{0.74}}&{\small \textbf{0.62}}&{\small \textbf{0.62}}&{\small \textbf{0.60}} \\
\hline
{\small $\boldsymbol{V}$} {\footnotesize \textbf{Med. \& High Spectral Res.}} &{\small \textbf{0.72}}&{\small \textbf{0.53}}&{\small \textbf{0.81}}&{\small \textbf{0.68}}&{\small \textbf{0.67}}&{\small \textbf{0.65}} \\
\hline
\end{tabular}
\end{center}
\end{table}

\subsection{Transmission and number of photon}
The number of photons produced by the source and collected by one telescope is :
\begin{equation}
n_*=T_*.S_{\rm tel}.\Delta\lambda.\Delta t.10^{-0.4m}.\Phi_0.\frac{\lambda}{hc},
\end{equation}
with the following parameters:
\begin{itemize}
\item $\Delta\lambda$ : spectral channel bandwidth.
\item $\Delta t$ : detector exposure time.
\item $S_{\rm tel}$ : spectral bandwidth.
\item $m$ : magnitude of the source.
\item $\Phi_0$ : reference flux at zero magnitude.
\item $T_*$ : overall transmission for a stellar source.
\end{itemize}
The number of photons of the source is then
$n^{\rm I}=2n_*/3$ in the interferometric channel, and $n^{\rm P} =n_*/3$ the in the photometric channels. 
The number of photons from the thermal background photon collected by one telescope is:
\begin{equation}
n_{\rm b}=T_{\rm B}.\epsilon_{\rm B}.S_{\rm tel}.\Delta\Omega.\Delta\lambda.\Delta t.B(\lambda).\frac{\lambda}{hc},
\end{equation}
with the following parameters:
\begin{itemize}
\item $T_{\rm B}$ : overall transmission for the thermal background.
\item $\epsilon_{\rm B}$ : global emissivity of the optical train.
\item $\Delta \Omega$ : solid angle of the interferometric field of view limited by the spatial filtering.
\item $B(\lambda)$ : blackbody intensity spectrum (Planck's function).
\end{itemize}
Also, $n_{\rm b}^{\rm I}=2n_{\rm b}/3$ and $n_{\rm b}^{\rm P}=n_{\rm b}/3$. The values of $T_*$, $T_{\rm B}$ and $\epsilon_{\rm B}$ are shown in Table~\ref{tab:trans}, and explicited in detail in Lagarde et al. (2012)\cite{2012SPIE.8445E..2JL}. 

\begin{table}[ht]
\caption{Estimated values for $T_*$, $T_{\rm B}$, and $\epsilon_{\rm B}$. 
} 
\label{tab:trans}
\begin{center}       
\begin{tabular}{ccccccc} 
\hline
\multirow{2}{*}{\textbf{}}&\multicolumn{2}{c}{\textbf{{\itshape L }band}} & \multicolumn{2}{c}{\textbf{{\itshape M} band}}& \multicolumn{2}{c}{\textbf{{\itshape N} band}}\\
\hline
&\textbf{AT}&\textbf{UT}&\textbf{AT}&\textbf{UT}&\textbf{AT}&\textbf{UT}\\
\hline
$T_*$&0.0085&0.007&0.0075&0.0065&0.04&0.038 \\
\hline
$T_{\rm B}$ &0.12&0.12&0.12&0.12&0.28&0.28 \\
\hline
$\epsilon_{\rm B}$ &0.78&0.80&0.85&0.85&0.79&0.80 \\
\hline
\end{tabular}
\end{center}
\end{table}

\subsection{Detector parameters}
The detector parameters used to calculate the MATISSE performance are:
\begin{itemize}
\item Detector readout noise: RON = 300e- in {\itshape N} band, and RON = 15e- in {\itshape L} and {\itshape M} bands.
\item Number of pixels read in the interferometric channel, $n_{\rm P}'$, and in a photometric channel, $n_{\rm P}''$ .A finite window of the detector is read (Sect. 3.1). The size of the window is typically $4\lambda/D$ in the spatial direction. The width of a spectral channel is $2\lambda/D$ in {\itshape N} band and $2\lambda/D$ in {\itshape L} band. In {\itshape N} band, the sampling is $3\times72$ pixels per $\lambda/D$ in the interferometric channel at $8 \mu$m. It leads to read 2977
pixels for the central wavelength (10.5~$\mu$m). In {\itshape L} band, the sampling is $3\times72$ pixels per $\lambda/D$ in the interferometric channel at 3.2~$\mu$m. It leads to read 1550 pixels for the central wavelength (3.5~$\mu$m). For the photometric channels, $n_{\rm P}''$ is 6 times lower.
\item Detector quantum efficiency: $Q.E$ = 50\% in all bands.
\item Detector readout mode: “correlated double sampling”.
\item Detector readout speed: the Pixel Clock Frequency is 1~MHz in {\itshape N} band, and 100 kHz in {\itshape L}  and {\itshape M} bands.
\item Detector Full Well Capacity: $FWC=0.6 10^{-6}$~e- in {\itshape N} band, and $FWC = 100000$~e- in {\itshape L} and {\itshape M} bands.
\item Detector pixel crosstalk (spatial direction): 10\% in {\itshape N} band, and 3.5\% in {\itshape L} and {\itshape M} bands.
\end{itemize}

\section{Performance}
The MATISSE performances are expressed in terms of limiting magnitude, and accuracy on the interferometric observables (visibility, closure phase, differential phase).

\subsection{Limiting magnitude} 
The limiting magnitude given here are the limiting magnitudes for coherent data processing. That is, each interferogram contains enough phase - i.e. piston - information, to allow coherent addition of the Fourier interferograms. Our requirement on the fringes detection is $SNR_{\rm C}=3$ after the integration of all the spectral channels read by the detector and integration of the frames acquired in one piston coherence time. We assumed a typical coherence time of 50 ms for the {\itshape L} and {\itshape M} bands, and 300 ms for the {\itshape N} band. In {\itshape N} band, this is made by recording several 30~ms frames at low spectral resolution (due to detector saturation with the thermal background) or 1 frame at medium resolution (R=220). As the criterion of $SNR_{\rm C}=3$ is obtained after integrating over all spectral channels in the band and all frames in one piston coherence time, the limiting magnitude is similar for the both spectral resolutions.
Table \ref{tab:magnitude} gives the limiting magnitudes for all the spectral bands of MATISSE in 'SiPhot' mode without fringe tracking. The reachable
limiting magnitude in medium and high spectral resolutions in {\itshape L} and {\itshape M} bands is limited by the fact that only a small part of the spectrum can be read by the detector (see readout speed in Sect. 5.3). This is not the case in {\itshape N} band. In addition, for
the longest wavelengths, for instance in {\itshape N} band, the measurements are photon noise dependant and not detector noise dependant.\\
\begin{table}[ht]
\caption{Limiting correlated magnitudes of MATISSE in 4T SiPhot mode. 
} 
\label{tab:magnitude}
\begin{center}       
\begin{tabular}{ccccc} 
\hline
&\multicolumn{2}{c}{\textbf{AT}}&\multicolumn{2}{c}{\textbf{UT}}\\
\hline
&$\boldsymbol{R=30}$&$\boldsymbol{R=500}$&$\boldsymbol{R=30}$&$\boldsymbol{R=500}$\\
\hline
\textbf{{\itshape L} band}&{\small 2.85 Jy (L=5)}&{\small 32.8 Jy (L=2.35)}&{\small 0.26 Jy (L=7.6)}&{\small 3 Jy (L=4.95)} \\
\hline
\textbf{{\itshape M} band}&{\small 7.6 Jy (M=3.45)}&{\small 60.4 Jy (M=1.2)}&{\small 0.58 Jy (M=6.25)}&{\small 4.4 Jy (M=4.05)} \\
\hline
&$\boldsymbol{R=30}$&$\boldsymbol{R=220}$&$\boldsymbol{R=30}$&$\boldsymbol{R=220}$\\
\hline
{\textbf{{\itshape N} band}}&{\small 14.6 Jy (N=1)}&{\small 14.6 Jy (N=1)}&{\small 0.9 Jy (N=4)}&{\small 0.9 Jy (N=4)} \\
\hline
\end{tabular}
\end{center}
\end{table}
The use of a fringe tracker allows an increase of the $DIT$ and the observation time. This has two impacts:
\begin{itemize}
\item It increases the number of spectral channels that can be read by the detector.
\item It decreases the detector noise contribution with respects to the photon noise.
\end{itemize}
For instance, with fringe tracking, a $DIT$ lower than 250~ms in {\itshape L} and {\itshape M} bands at medium spectral resolution and a $DIT$ of about 350~ms in {\itshape L} band at high spectral resolution are sufficient to obtain the same limiting
magnitudes that the ones estimated for the low resolution without fringe tracking (see table~\ref{tab:magnitude}).
Table~\ref{tab:magnitude_ft} gives the limiting magnitudes in {\itshape L} band in 4T SiPhot mode with a
polarization selection for two cases: if no fringe tracking is used ($DIT=50$~ms) and when a fringe tracking is performed ($DIT=300$~ms). Table~\ref{tab:magnitude_ft} also gives the limiting magnitudes in {\itshape N} band in 4T SiPhot mode without and with a fringe tracking. In {\itshape N} band, only the observation time can be increased since detector saturation due to the thermal background prevents from recording frames longer than about 30~ms. 
\begin{table}[h]
\caption{Expected sensitivity performances of MATISSE in SiPhot mode, without and with an external fringe tracker. Low spectral resolution ($R=30$) is considered. The integration and observation times with the fringe tracker are typical values, given as a guide, and may be increased to reach higher sensitivities.  
} 
\label{tab:magnitude_ft}
\begin{center}       
\begin{tabular}{ccccc} 
\hline
&\multicolumn{2}{c}{\textbf{{\itshape L }band sensitivity}} & \multicolumn{2}{c}{\textbf{{\itshape N }band sensitivity}} \\
\hline
&{\small Without FT} & {\small With FT}&{\small Without FT} &{\small With FT} \\
&&{\footnotesize (DIT=300ms)}&&{\footnotesize (Obs=10s)} \\
\hline
\textbf{AT} & {\small 2.95 Jy ({\itshape L}=5)} & {\small 0.55 Jy ({\itshape L}=6.8)}& {\small 14.6 Jy ({\itshape N}=1)} & {\small 2.1 Jy ({\itshape N}=3.1)} \\
\textbf{UT} & {\small 0.26 Jy ({\itshape L}=7.6)} & {\small 0.05 Jy ({\itshape L}=9.5)}& {\small 0.9 Jy ({\itshape N}=4)} & {\small 0.12 Jy ({\itshape N}=6.2)} \\
\hline 
\end{tabular}
\end{center}
\end{table}
We have a typical gain of 2.5 magnitudes offered by the use of a fringe tracker. In addition, this gain
can be higher with a higher detector integration time.

\subsection{Accuracies on calibrated observables}
Here we give the accuracies on the absolute visibility, the closure phase, and the differential phase, taking into account the fundamental noises and the calibration procedure. The conditions are the following:
\begin{itemize}
\item SiPhot mode is considered.
\item The estimates are given per spectral channel for a chopping frequency of 0.8 Hz corresponding to the following
sequences: 600 ms target observation, 25 ms chopping (telescope pointing, OA and FT close loops), 600 ms sky.
\item The performance is computed for an observation of 15 minutes.
\item In {\itshape L} and {\itshape M} bands, the Detector Integration Time is $DIT=30$~ms without fringe tracking and the frame time is 600 ms with fringe tracking.
\item In {\itshape N} band, the Detector Integration Time is $DIT=30$~ms.
\end{itemize}
The errors on visibilities and phases are also affected by the transfer function variation between the observations of the science object and the calibrator (typically 30 minutes). The main contributions to this transfer function variation are detailed and quantified in Lagarde et al. (2012)\cite{2012SPIE.8445E..2JL}. We summarize them hereafter :
\begin{itemize}
\item a change in the image and pupil position: the effect is negligible thanks to the spatial filtering.
\item An OPD variation, which can actually be calibrated.
\item A change in the image overlap: this effect concerns the differential drift between the
superimposition of the images of the spatial filters.
\item A Strehl ratio variation between the two observations, which can be limited by the spatial filtering.
\item The OPD jitter between the two observations. This OPD jitter variation has no effect on the closure phase.
\end{itemize}
All these contributions have not the same impacts on the measurements. The absolute visibility and the closure phase are
directly affected by the absolute value of the change. The differential visibility and the differential phase are only affected by the differential chromatic effects of this change.\\
In addition, the estimation of the closure phase could be degraded by several detection effects, a crosstalk between fringe peaks, window effects, detector drifts, and parasitic lights within the instrument. So, the fundamental precision limits could be difficult to achieve in practice \cite{Petrov2010}. This error will be quantified during the commissioning of MATISSE. 
The estimation of the differential phase could also be degraded by several instrumental and atmospheric parameters:
\begin{itemize}
\item The atmosphere+tunnel contribution to the measured differential phase on
baseline $ij$. This is the dominant variable chromatic OPD term.
\item The VLTI optics contribution to the measured differential phase on baseline $ij$. This refers to all instrumental terms produced before the beam commuting device (BCD).
\item The instrument contribution to the measured differential phase on baseline $ij$.
\end{itemize}
As for the closure phase, this error will be quantified during the commissioning of MATISSE. The Chromatic OPD term will be computed in MATISSE using ambient measurements and pre-defined dispersion laws from \cite{1996ApOpt..35.1566C,2007JOptA...9..470M}. In addition, the wide spectral
coverage of MATISSE could allow a better fit of the chromatic OPD, as it was attempted by Matter et al. (2010) \cite{2010A&A...515A..69M} with AMBER and MIDI.
Table~\ref{tab:accuracy} gives the accuracies on visibilities and on phases (closure and differential) for a 20 Jy source taking into account the fundamental noises, the transfer function variation, and the additional errors on the phases due to the detection and chromatic effects. The use of a fringe tracking is crucial for faint sources and for
observations with ATs or at medium (and high) spectral resolution.

\begin{table}[ht]
\caption{Calibrated measurement accuracies for a 20 Jy source. FT = Fringe Tracker.
} 
\label{tab:accuracy}
\begin{center}       
\begin{tabular}{lcccccccc} 
\hline
&\multicolumn{4}{c}{$\boldsymbol{R=30}$} & \multicolumn{4}{c}{$\boldsymbol{R=500}$}\\
\cline{2-9}
\textbf{{\itshape L} band}&\multicolumn{2}{c}{\textbf{AT}} & \multicolumn{2}{c}{\textbf{UT}} & \multicolumn{2}{c}{\textbf{AT}}  & \multicolumn{2}{c}{\textbf{UT}}\\
\cline{2-9}
&\textbf{No FT}&\textbf{FT}&\textbf{No FT}&\textbf{FT}&\textbf{No FT}&\textbf{FT}&\textbf{No FT}&\textbf{FT}\\
\hline
{\small \textbf{Absolute visibility}}&{\small 1.7\%}&{\small 1.6\%}&{\small 2.3\%}&{\small 2.5\%}&/&{\small 2.3\%}&{\small 2.3\%}&{\small 2.5\%}\\
\hline
{\small $\boldsymbol{\sigma_{\psi}}$~\textbf{(mrad)}}&{\small 20.3}&{\small 20}&{\small 20}&{\small 20}&/&{\small 25.1}&{\small 20.3}&{\small 20.1}\\
\hline
{\small $\boldsymbol{\sigma_{\phi}}$~\textbf{(mrad)}}&{\small 19.3}&{\small 18.8}&{\small 22.2}&{\small 23}&/&{\small 20.8}&{\small 22.3}&{\small 23}\\
\hline
\hline
&\multicolumn{4}{c}{$\boldsymbol{R=30}$} & \multicolumn{4}{c}{$\boldsymbol{R=500}$}\\
\cline{2-9}
\textbf{{\itshape M} band}&\multicolumn{2}{c}{{\small \textbf{AT}}} & \multicolumn{2}{c}{{\small \textbf{UT}}} & \multicolumn{2}{c}{{\small \textbf{AT}}}  & \multicolumn{2}{c}{{\small \textbf{UT}}}\\
\cline{2-9}
&{\small \textbf{No FT}}&{\small \textbf{FT}}&{\small \textbf{No FT}}&{\small \textbf{FT}}&{\small \textbf{No FT}}&{\small \textbf{FT}}&{\small \textbf{No FT}}&{\small \textbf{FT}}\\
\hline
{\small \textbf{Absolute visibility}}&{\small 1.3\%}&{\small 1.1\%}&{\small 1.5\%}&{\small 1.7\%}&/&{\small 2.9\%}&{\small 1.6\%}&{\small 1.7\%}\\
\hline
{\small $\boldsymbol{\sigma_{\psi}}$~\textbf{(mrad)}}&{\small 15}&{\small 14.1}&{\small 13.7}&{\small 13.7}&/&{\small 29}&{\small 14.3}&{\small 13.8}\\
\hline
{\small $\boldsymbol{\sigma_{\phi}}$~\textbf{(mrad)}}&{\small 17.3}&{\small 16.9}&{\small 18.5}&{\small 19.1}&/&{\small 22.4}&{\small 18.6}&{\small 19.1}\\
\hline
\hline
&\multicolumn{4}{c}{$\boldsymbol{R=30}$} & \multicolumn{4}{c}{$\boldsymbol{R=220}$}\\
\cline{2-9}
\textbf{{\itshape N} band}&\multicolumn{2}{c}{{\small \textbf{AT}}} & \multicolumn{2}{c}{{\small \textbf{UT}}} & \multicolumn{2}{c}{{\small \textbf{AT}}}  & \multicolumn{2}{c}{{\small \textbf{UT}}}\\
\cline{2-9}
&{\small \textbf{No FT}}&{\small \textbf{FT}}&{\small \textbf{No FT}}&{\small \textbf{FT}}&{\small \textbf{No FT}}&{\small \textbf{FT}}&{\small \textbf{No FT}}&{\small \textbf{FT}}\\
\hline
{\small \textbf{Absolute visibility}}&{\small 8.6\%}&{\small 8.3\%}&{\small 2.8\%}&{\small 1.2\%}&{\small 27.1\%}&{\small 25.8\%}&{\small 3.2\%}&{\small 1.7\%}\\
\hline
{\small $\boldsymbol{\sigma_{\psi}}$~\textbf{(mrad)}}&{\small 28.2}&{\small 22.5}&{\small 13.6}&{\small 13.6}&{\small 145.5}&{\small 106.3}&{\small 16.3}&{\small 14.7}\\
\hline
{\small $\boldsymbol{\sigma_{\phi}}$~\textbf{(mrad)}}&{\small 26.1}&{\small 19.7}&{\small 24.9}&{\small 16.8}&{\small 86.5}&{\small 63.1}&{\small 25.4}&{\small 17.1}\\
\hline
\end{tabular}
\end{center}
\end{table}

\section{Upgrades at the VLTI}
The full exploitation of the potential of MATISSE requires instrumental upgrades at the VLTI. This concerns in particular the implementation of adaptive optics on the ATs, and, most important, the availability of an external fringe tracker for MATISSE. An external fringe tracker will improve the sensitivity, accuracy and spectroscopic capabilities of MATISSE, with a direct impact on the scientific potential of the instrument for the study of the protoplanetary disks in particular:
\begin{itemize}
\item reaching sensitivities below a few tens of mJy in the mid-infrared would allow MATISSE to access a statistically significant number of sources (a few hundred) of different ages and masses, for both model fitting and image reconstruction studies. Probing different age and mass regimes is fundamental to understand the disks evolution scheme down to the faintest sources represented by the solar-mass young stars (T Tauri stars). 
\item High accuracy phase measurements ($\sim 1$~millirad) will be important for closure-phase imaging in the {\itshape L}, {\itshape M}  and {\itshape N} bands, which will especially enable the non-ambiguous detection and study of complex au-scale structures and asymmetries in the inner region of disks.  
\item With the longer integration times allowed by an external fringe tracking device, medium and high spectral resolution interferometry will be made feasible over the full range of the spectral bands of MATISSE.
\end{itemize}
Two upgrades are planned at the VLTI : NAOMI, the Adaptative Optics system for the ATs, and GRA4MAT. The GRA4MAT project will enable the use of the GRAVITY Fringe Tracker (FT) to track fringes in {\itshape K} band while measuring the interferometric observables in {\itshape L\&M} and {\itshape N} bands with MATISSE. In this process, the main limiting factor is the {\itshape K} band sensitivity of the GRAVITY FT. This sensitivity has been estimated to $K_{\rm corr}\simeq7$ during the GRAVITY commissioning with the ATs, and is expected to be $K_{\rm corr}\simeq10$ with the UTs.
Compared to these values, we can expect a sensitivity gain of 0.8 to 1 magnitude in the case of GRA4MAT, where all the {\itshape K} band flux will be sent to the fringe tracker. This is not the case for GRAVITY alone, where only 50\% of the flux is available for fringe tracking.
Then, provided fringes can be tracked in {\itshape K} band, we can expect a minimum gain of about 2 magnitudes at the MATISSE wavelengths (see Table \ref{tab:magnitude_ft}). 
The current plan is to have GRA4MAT commissioned late 2018, at the end of the MATISSE commissioning.   

\section{MATISSE Simulation tools}
The specifications and expected performances summarized in this article were mostly taken from the first performance analysis made for the Final Design Review in 2012. The performance analysis is thus still on-going since it depends on aspects that are not fully set yet such as the SNR equations and the noise model of MATISSE or various instrumental parameters values (instrumental throughput, calibration errors, ...). These elements will be better defined and the MATISSE performances validated during the test phase in Nice and the commissioning on sky in Paranal. This is essential for simulating the instrument and thus assessing the feasibility of the future MATISSE observing programs.
In this context, a reference MATISSE simulator has been developed at OCA (Observatoire de la C\^ote d'Azur). It is an internal tool based on the 'MATISSE Performance Analysis Report' ESO document. Close to the instrument, this tool will be regularly updated during the test and commissioning phases, in terms of sources of noise, $SNR$ equations, and instrumental parameters. This tool serves as a reference for the simulation tools that will be made available to the future MATISSE users. This will include :
\begin{itemize}
\item a contractual Exposure Time Calculator (ETC) delivered by ESO. Its objective will be to check the feasibility of MATISSE observations by providing various outputs such as $SNR$ and error estimates for a given observing time, or required observing times for a given $SNR$ and error level. This tool will be used through a web-based user-friendly interface based on the existing GRAVITY ETC interface.
\item A module and noise model simulating MATISSE, which is being implemented in the ASPRO2 software. ASPRO2 is a complete observation preparation tool developed and maintained by the JMMC, which allows to prepare interferometric observations with the VLTI or other interferometers. We refer to Bourg\`es et al. (2016, this volume) for more details on the implementation of the noise model of MATISSE into ASPRO2. 
\end{itemize}

\section{Conclusion}
MATISSE is a ESO contractual instrument for which technical specifications were defined to address a rich astrophysical program\cite{Lopez13}, in particular the observation of the inner regions of protoplanetary disks, and the active galactic nuclei. The current MATISSE performance analysis shows that the expected performances are even better than the specifications, making
unmatched the expectations of the future instrument. All the sub-systems of MATISSE are now fully integrated at the Observatoire de la Cote d'Azur. The test phase has started in June 2016
and will end in March 2017. Once the test phase has been completed, the ESO review called Preliminary
Acceptance Europe will be conducted between March 2017 and July 2017. It
will give the green light for the shipping of the instrument to Paranal
and its integration in September 2017. After the period called Assembly,
Integration and Verification, the Commissioning will take place early
2018. If such a schedule is maintained the first observing proposals can
be submitted for September 2018. This will lead to the first science in 2019. 

 
\bibliography{report} 
\bibliographystyle{spiebib} 

\end{document}